\documentclass[12pt,a4paper,dvips]{article}
\usepackage{a4p}
\usepackage{cite,mcite}
\usepackage{graphicx}
\usepackage{epsfig}
\usepackage{l3_title,ifthen}
\usepackage{physics,Lep}
\usepackage{a4p}
\date{September 20, 1999}
\preprint{99-127}
\def\dm        {\ensuremath{\Delta M}}
\def\susy#1{\ensuremath{\tilde{\mathrm{#1}}}}%
\def\slepton   #1{\ensuremath{\susy{\ell}^{#1}}}
\def\selectron #1{\ensuremath{\susy{e}^{#1}}}

\def\stau      #1{\ensuremath{\susy{\tau}^{#1}}}
\def\sneutrino {\susy{\nu}}
\def\chargino  #1{\ensuremath{\susy{\chi}_1^{#1}}}
\def\charginos  #1{\ensuremath{\susy{\chi}_{1,2}^{#1}}}
\def\neutralino#1{\ensuremath{\susy{\chi}_{#1}^0}}

\def\Zstar     {\ensuremath{\mathrm{Z^\ast}}}
\def\Wstar     {\ensuremath{\mathrm{W^\ast}}}
\def\EV30{\ensuremath{E_{\mathrm{v30}}}}%
\def\Ebar{\ensuremath{E\hspace{-.23cm}/\hspace{+.01cm}}}
\def\EM25{\ensuremath{\Ebar_{25}}}
\def\EMF25{\ensuremath{\Ebar^{\perp}_{25}}}
\def\ECM60{\ensuremath{\Ebar^b_{60}}}

\def\TKM25{\ensuremath{N_{tk}^{25}}}

\def\DELTAM{\ensuremath{\Delta{M}}}%

\def\simge{\ \lower -2.5pt\hbox{$>$} \hskip-8pt \lower 2.5pt \hbox{$\sim$}\ }
\def\simle{\ \lower -2.5pt\hbox{$<$} \hskip-8pt \lower 2.5pt \hbox{$\sim$}\ }

\newlength{\capindent}
\newlength{\capwidth}
\newlength{\figwidth}
\newcommand{\icaption}[2][!*!,!]{\hspace*{\capindent}%
  \begin{minipage}{\capwidth}
    \ifthenelse{\equal{#1}{!*!,!}}%
      {\caption{#2}}%
      {\caption[#1]{#2}}
  \end{minipage}}
\begin{document}
\begin{titlepage}
\title{Search for  Charginos and  Neutralinos   
\\ \boldmath in e$^+$e$^-$ collisions   at $\sqrt{s} = 189 \gev{}$}
\author{The L3 collaboration}
\vspace{1.5cm}

%
%
\begin{abstract}
We report the result of a search for 
charginos and neutralinos, in \epem\, 
collisions at 189 \gev{} centre-of-mass energy at LEP.
No evidence for such particles is found in a data sample of 176 pb$^{-1}$.
Improved upper limits for these particles
are set on the production cross sections.
New exclusion contours in the parameter space of the
Minimal Supersymmetric Standard Model are derived, 
as well as new lower limits on the masses of these
supersymmetric particles.
Under the assumptions of common gaugino and scalar masses at the GUT
scale, we set an absolute lower limit on the mass of the lightest
neutralino of 32.5 \gev{} and on the mass of the lightest
chargino of 67.7 \gev{}.
\end{abstract}

\submitted


\end{titlepage}

%
%
%
\section{Introduction}

One of the main goals of the LEP experiments is to search for new
particles predicted by theories beyond the Standard
Model. In this letter we report on searches for 
unstable  charginos and neutralinos.
These particles are predicted by supersymmetric theories (SUSY)
\cite{susy}. In SUSY theories with minimal particle
content (MSSM) \cite{mssm}, in addition to the ordinary particles, there is
a supersymmetric spectrum of particles with spins which differ by one
half with respect to their Standard Model partners.

Charginos (\charginos{\pm}), the supersymmetric partners of \Wpm\, 
and \Hpm, are pair produced via $s$-channel 
$\gamma/\mathrm{Z}$ exchange. The production cross section can be
reduced by an order of magnitude when the $t$-channel scalar neutrino
(\sneutrino) exchange is important.
Neutralinos, the supersymmetric partners of Z, \gam, and neutral
Higgs bosons, are pair produced  
$\ee \rightarrow \neutralino{i}\neutralino{j}~ (i,j=1, \ldots ,4$;
ordered by their masses) via $s$-channel $\mathrm{Z}$ exchange and their
 production cross section can be enhanced by $t$-channel exchange of a 
scalar electron (\selectron{\pm}). 

Short-lived supersymmetric particles are expected
in R-parity conserving SUSY models. 
The R-parity is a quantum number which 
distinguishes ordinary particles from supersymmetric particles.
If R-parity is conserved, supersymmetric particles are 
pair-produced and the lightest supersymmetric particle, 
the lightest neutralino, \neutralino{1}, is stable. The neutralino is weakly-interacting 
and escapes detection.
In this letter we assume R-parity conservation, which implies 
that the decay chain of supersymmetric particles always contain,
besides standard particles, two invisible neutralinos
causing the missing energy signature.

When the masses of the scalar leptons and the charged Higgs bosons (\Hpm) are very large,
the \chargino{\pm} decays via \Wstar:  
$\chargino{\pm} \rightarrow \neutralino{1}\Wstar
                \rightarrow \neutralino{1}\, f\bar{f}^\prime$.
If the \slepton{\pm} and \sneutrino\, masses are comparable to 
$M_{\W}$ the chargino also 
decays via virtual scalar lepton or scalar neutrino 
and the leptonic branching fraction is enhanced.
Finally for \slepton{\pm} and \sneutrino\, lighter than the chargino, 
the decay modes 
$\chargino{\pm} \rightarrow \slepton{\pm} \nu$ or
$\chargino{\pm} \rightarrow \sneutrino \ell^\pm$ become dominant. 
When the masses of the neutral SUSY Higgs bosons (\ho, \Ao) and of
the scalar leptons are very large, the heavier neutralinos 
($\neutralino{j},\ j \ge 2$) decay via \Zstar:
$\neutralino{j} \rightarrow \neutralino{k}\Zstar
                \rightarrow \neutralino{k}\, f\bar{f}$  with $k < j$.
For a chargino lighter than neutralinos, the latter decay via
$\Wstar$ such as $\neutralino{j} \rightarrow \chargino{\pm} f\bar{f}^\prime$.
If the scalar lepton masses are comparable to the Z mass,
the neutralino decays also via a virtual scalar lepton, enhancing the leptonic 
branching fraction.
Finally, for \sneutrino\, and \slepton{\pm} lighter than neutralinos the
two-body decays 
$\neutralino{j} \rightarrow \slepton{\pm} \ell^\mp$ or
$\neutralino{j} \rightarrow \sneutrino \nu$ (j $\ge 2$)
become dominant.    
The radiative decays $\neutralino{j} \rightarrow \neutralino{k} \gamma $ 
are also possible via higher-order diagrams.

Previous results on chargino and neutralino searches have been 
reported by L3~\cite{reflep1.5,susy_96,susy_97} 
and other LEP experiments~\cite{charlep98}.
In this letter, new limits are presented on chargino 
and neutralino production cross sections.
These experimental results are interpreted in 
the framework of the constrained MSSM. 
Within these models lower limits on the masses of 
supersymmetric particles are derived.
For these limits present experimental results are combined 
with those obtained previously by L3 
at the Z peak \cite{oldsusyl3} and at energies up to
$\sqrt{s} = 183 \gev{}$ \cite{reflep1.5,susy_96,susy_97,nota99slep}.

\section{Data Sample and Simulation} \label{dtsmcs}

We present the analysis of data collected by
the L3 detector \cite{l3-detector} in 1998, 
corresponding to an integrated luminosity of 176.3 pb$^{-1}$ at an
average centre-of-mass energy of 188.6 \gev{}, denoted hereafter as 
$\sqrt{s} = 189 \gev{}$.

Standard Model reactions are simulated with the following 
Monte Carlo generators:
{\tt PYTHIA}~\cite{PYTHIA} for 
  $\ee \rightarrow \mathrm{q\bar{q}}$,
  $\ee \rightarrow \mathrm{Z}\,\ee$ and 
  $\ee \rightarrow \gamma\!/\mathrm{Z}\,\gamma\!/\mathrm{Z} $;
{\tt EXCALIBUR}~\cite{EXCALIBUR} for
   $\ee \rightarrow \mathrm{W^\pm\, e^\mp \nu}$;
{\tt KORALZ}~\cite{KORALZ} for
   $\ee \rightarrow \mu^+\mu^-$ and
   $\ee \rightarrow \tau^+\tau^-$;
{\tt BHWIDE}~\cite{BHWIDE} for 
   $\ee \rightarrow \ee$;
{\tt KORALW}~\cite{KORALW} for
   $\ee \rightarrow \mathrm{W^+ W^-}$;
two-photon interaction processes have been simulated using 
{\tt DIAG36}~\cite{DIAG} ($\ee \rightarrow \ee \ell^+\ell^-$) and
{\tt PHOJET}~\cite{PHOJET} ($\ee \rightarrow \ee\, \mathrm{hadrons}$), 
requiring at least 3 \gev{} for the invariant mass of the two-photon system.
The number of simulated events for each background process is 
equivalent to more than 100 times the statistics of the collected 
data sample except for two-photon interactions for which it is more 
than two times the data statistics.

Signal events are generated with the Monte Carlo program 
{\tt SUSYGEN}~\cite{susygen2.2}, for masses of SUSY particles 
($M_{\rm SUSY}$) ranging from 45 \gev{} up to the kinematic limit and for
$\DELTAM$ values 
($\dm = M_{\rm SUSY} - M_{\neutralino{1}}$) between 3 \gev{} and
$M_{\rm SUSY}-1 \GeV{}$. 
The explicit two-body decay branching ratios for charginos 
$\chargino{\pm} \rightarrow \sneutrino \ell^\pm, \slepton{\pm} \nu $ 
or  $\neutralino{2,3,4} \rightarrow \sneutrino \nu , \slepton{} \ell$
have been estimated with {\tt SUSYGEN}.

The detector response is simulated using the {\tt GEANT} 
package~\cite{geant}. It takes into account effects of energy loss,
multiple scattering and showering in the detector materials and
in the beam pipe. Hadronic interactions are simulated with the
{\tt GHEISHA} program~\cite{gheisha}. Time dependent inefficiencies
of the different subdetectors are also taken into account
in the simulation procedure.

%
%
\section{Analysis Procedure}

\subsection{Signal topologies and optimisation procedure}
\label{sec:optimization}

Besides the main characteristic of missing transverse momentum,
supersymmetric particle signals can be further specified according
to the number of leptons or the multiplicity of hadronic jets in the
final state. As mentioned in the introduction, chargino pair
production gives final states similar to \W\W\, production.
For neutralinos, we distinguish two classes of detectable processes: 
$\ee \rightarrow \neutralino{1}\neutralino{2}$ and 
$\ee \rightarrow \neutralino{2}\neutralino{2}$.
For these signals, final states are given by the Z branching ratios.
Both for charginos and neutralinos, the event energy
is directly related to \dm\, ($\dm = M_{\rm SUSY} - M_{\neutralino{1}}$).

We devise five types of selection criteria oriented to all decays 
of charginos, as follows:
at least two acoplanar leptons (e,$\mu$);  hadrons and at least one isolated lepton;
 at least two acoplanar taus;  hadrons and at least one isolated tau;
 purely hadronic final states with high multiplicity.
$\neutralino{2}\neutralino{2}$ production gives rise to final states very similar
to those of chargino pair production, even if with very different 
branching ratios. Hence, chargino selections based on these five topologies are also effective
to select $\neutralino{2}\neutralino{2}$ events.

The two-acoplanar-jets final state on the other hand deserves a
dedicated selection since it accounts for 70\% of the decays in
$\neutralino{1}\neutralino{2}$ events, and 28\% in
$\neutralino{2}\neutralino{2}$ events.

The signal topologies and the associated background sources depend strongly 
on \dm. Therefore all five selections are optimised separately for four
different \dm\, ranges:  
the very low \dm\, range at $3-5 \gev{}$, 
the low \dm\, range at $10-30 \gev{}$, 
the medium \dm\, range at $40-70 \gev{}$ and the high range
\dm\, at $80-94 \gev{}$.
In the low and very low \dm\, ranges, the expected topologies for the signal are 
characterised by a low multiplicity and a low visible energy, and
 the background is dominated by two-photon interactions.
For medium and high \dm\, ranges, the signal signatures are very similar to those of
W-pair production; in particular for $\dm > 80 \gev{}$ on-shell Ws are produced.

The cut values of each selection are {\it a priori} optimised using
Monte Carlo signal and background events. 
The optimisation procedure varies all cuts simultaneously to maximise the
signal efficiency and the background rejection.
In fact, the average limit ($\kappa^{-1}$) is minimised for an infinite  
number of tries, 
assuming only background contributions. This is expressed mathematically
by the following formula:
\begin{equation}  
\kappa=\epsilon/ \Sigma_{n=0}^{\infty} k(b)_{n} P(b,n)
\end{equation}
where $k(b)_{n}$ is the 95\% confidence level Bayesian upper limit, 
$P(b,n)$ is the Poisson distribution 
for $n$ events with an expected background of $b$ events, and 
$\epsilon$ is the signal efficiency.

\subsection{Event selection}

Lepton and photon identification, and isolation criteria in hadronic events
are unchanged compared to our previous analysis \cite{susy_96}.
The Durham algorithm~\cite{durham} is used for the clustering of
hadronic jets.

Events are first selected by requiring at least 3 \gev{} of visible energy
and 3 \gev{} of transverse momentum. Beam-gas events are rejected by
requiring the visible energy in a cone of $30^\circ$ 
around the beam
pipe ($E_{30^0}$) to be less than 90\% of the total and the missing momentum
vector to be at least $10^\circ$ away from the beam pipe.
Tagged two-photon interactions are rejected by requiring the sum of
the energies measured in the lead-scintillator
ring calorimeter and in the luminosity monitors~\cite{l3-detector}
to be less than 10 \GeV{}.
These two detectors cover the polar angle range 
$1.5^\circ < \theta <9^\circ$ on both sides of the interaction 
point.


\subsubsection{Leptonic final states} \label{sec:selection_lep}

For the pure leptonic final states, dedicated selections have been
optimised for the charginos, where the two leptons 
may have a different flavour. Those selections are very  
similar to the scalar lepton selections which are described in 
Reference \cite{slep99ref}. At the end, a combination of all
the leptonic selections, providing the optimal sensitivity, is done for
the chargino and the neutralino leptonic decays.

\subsubsection{Lepton plus hadrons final states} \label{sec:selection_lephad}

We select events with at least one isolated electron, muon or tau for which 
the energy, not associated to the lepton, in a cone of $30^\circ$ 
half-opening angle around its direction is less than 2 \GeV{}. 
The following quantities are defined:
the energy depositions
($E^{\perp}_{25}$ and $E_{25}$)
within $\pm25^\circ$ around the missing 
energy direction in the R--$\phi$ plane 
 or in space, respectively. 
We apply cuts on the number of tracks in the hadronic system
($N_{tk} - N_{lep}$) and the number of calorimetric clusters ($N_{cl}$).
Furthermore, cuts are applied on the missing energy direction isolation
($\theta_{miss}$ and $E^{\perp}_{25}$),
the total transverse momentum ($p_{\perp}$),
the energy of the isolated lepton ($E_{lep}$), 
the recoil mass ($M_{rec}$),
as well as on the acoplanarity angle between
the jet and the lepton. A cut is applied on the visible energy ($E_{vis}$)
and $E_{TTJL}$
which is defined as the absolute
value of the projection of the total momentum of the jet and the 
lepton onto the
direction perpendicular to the lepton-jet thrust computed in the R-$\phi$ 
plane. 
A cut on the invariant mass of the hadronic system ($M_{had}$) 
removes most of the WW background.

The cut values at $\sqrt{s} = 189 \gev{}$, are shown in Table~\ref{tab2} for the 
different \dm\ ranges.

\subsubsection{Purely hadronic final states} \label{sec:selection_hadrons}

The list of cuts used at $\sqrt{s} = 189 \gev{}$ is reported in Table~\ref{tab3} for the different 
\dm\ ranges. 
Again, we apply cuts on $N_{cl}$, $N_{tk}$, $p_{\perp}$, $E_{vis}$,
acollinearity and acoplanarity as well as on the missing energy 
polar angle ($\theta_{miss}$) and isolation ($E^{\perp}_{25}$, $E_{25}$).
The absolute value of the total momentum of the event 
along the beam line normalised to the visible energy ($p_\parallel$),
the recoil mass ($M_{rec}$) and the visible mass ($M_{vis}$) are also used
in the selections.

In all the selections, but the very low \dm,
a cut on the width of the two jets 
is applied. We define $y_{\perp}$ as the ratio between the scalar sum of the
particle momenta transverse to the jet direction and the jet energy.  
We require $y_{\perp}$ to be large in order to select four-jet-like events.
In the low \dm\ range a cut on the ratio $E_{TTJ}$/$p_\perp$ 
is applied. 
$E_{TTJ}$ is equivalent to $E_{TTJL}$ using the momenta and the directions of
the two jets.

\section{Results}

The results at $\sqrt{s} = 189 \gev{}$, for 
the eighteen chargino selections and the four neutralino
selections are shown in Table \ref{tab8}. The results for the very low and low \dm\, selections are shown together.
A good agreement between the expected background from Standard Model
processes and the selected data is observed.

The eighteen chargino selections
find 147 candidates in the data when expecting 148 events
from the Standard Model processes. 
In the low and
very low \dm\ regions
72 events are selected, 11 events in the medium \dm\, region and
67 events in the high \dm\, region. 
In the four neutralino selections
50 candidates are found 
whereas 48.1 events are expected
from the Standard Model processes, 
most of those events are selected by the 
low $\dm$ selections.

\begin{figure}
\begin{tabular}{cc}
\psfig{file=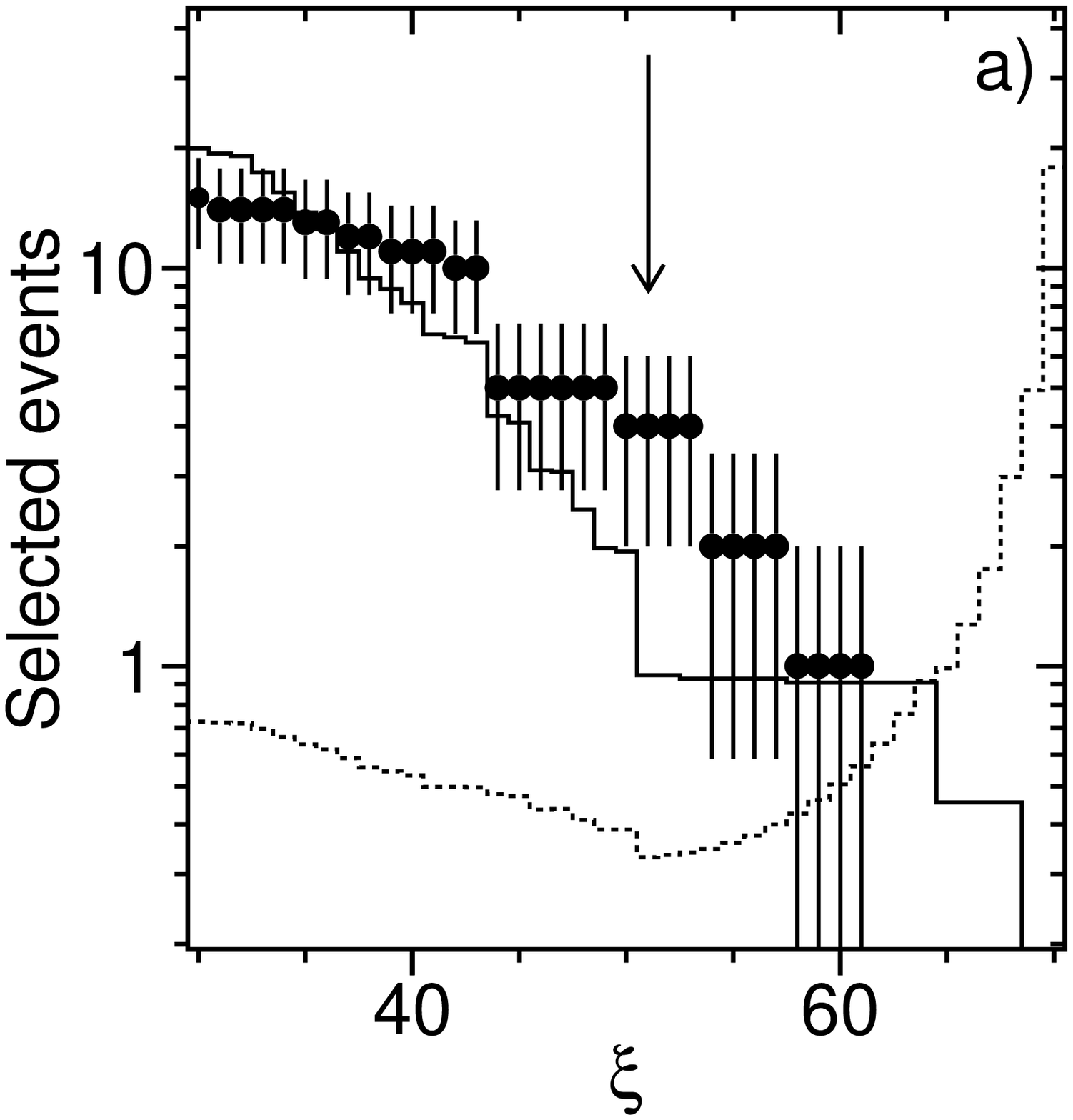,width=8.0cm} &
\psfig{file=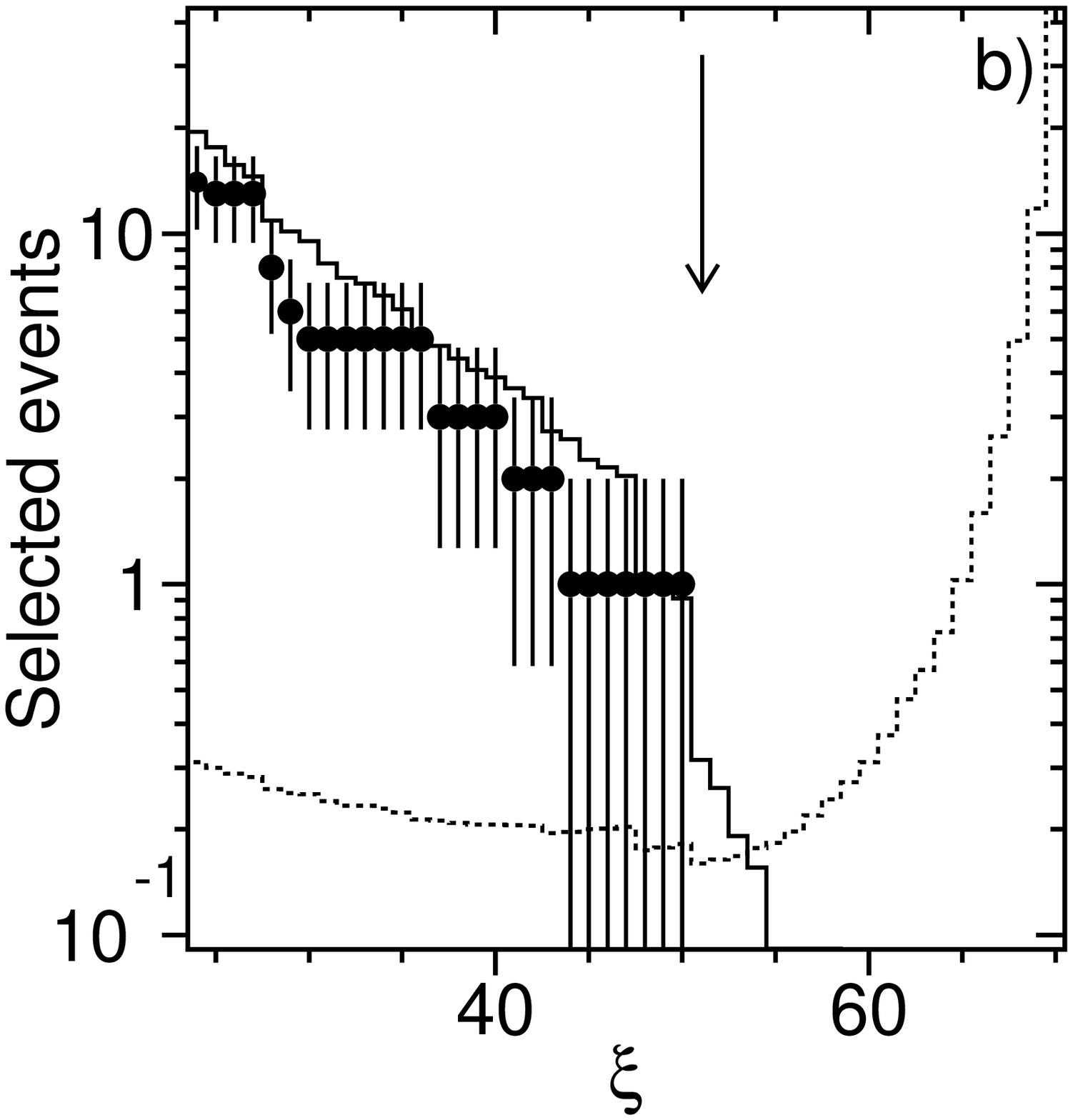,width=8.0cm} \\
\psfig{file=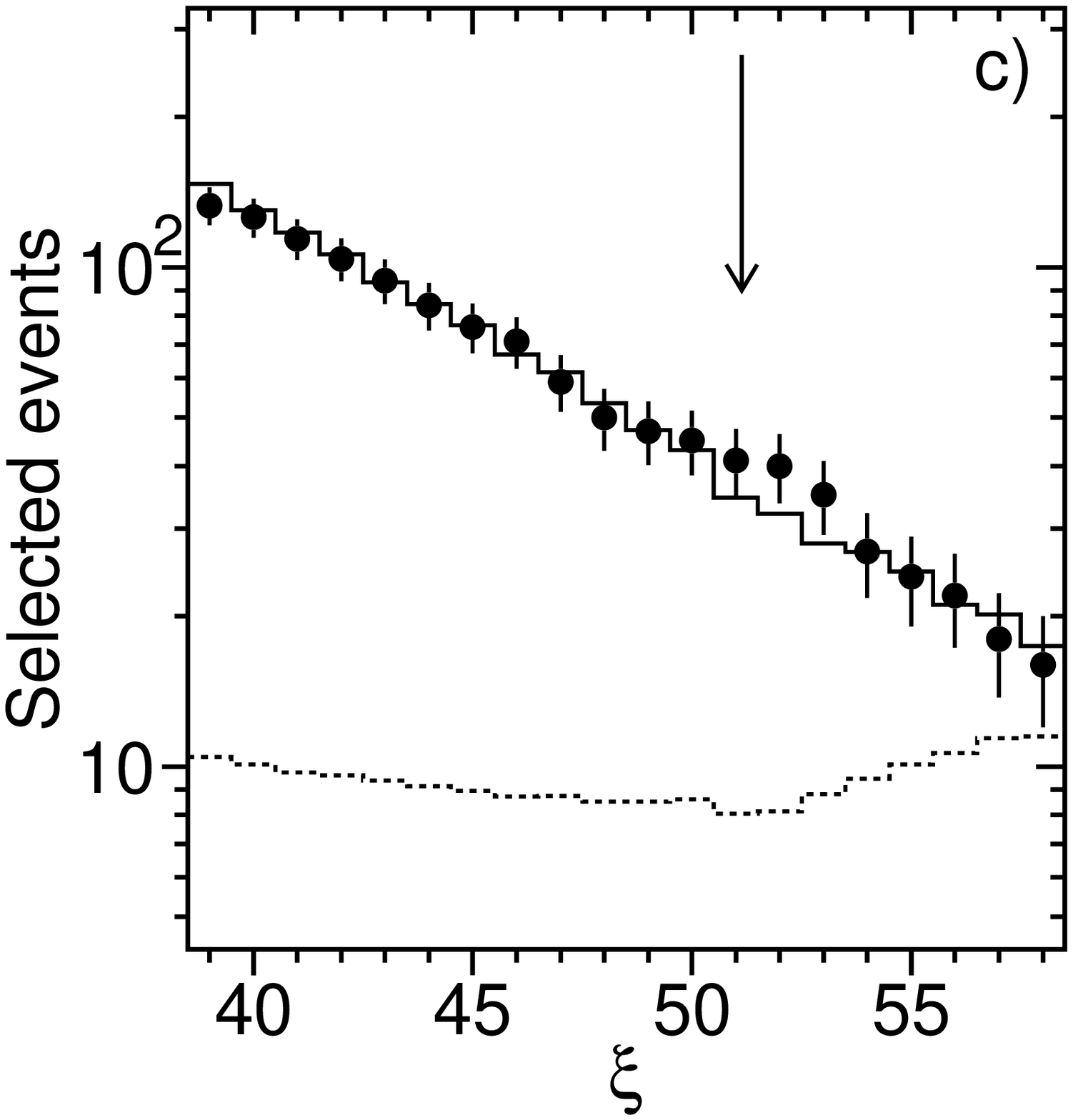,width=8.0cm} &
\psfig{file=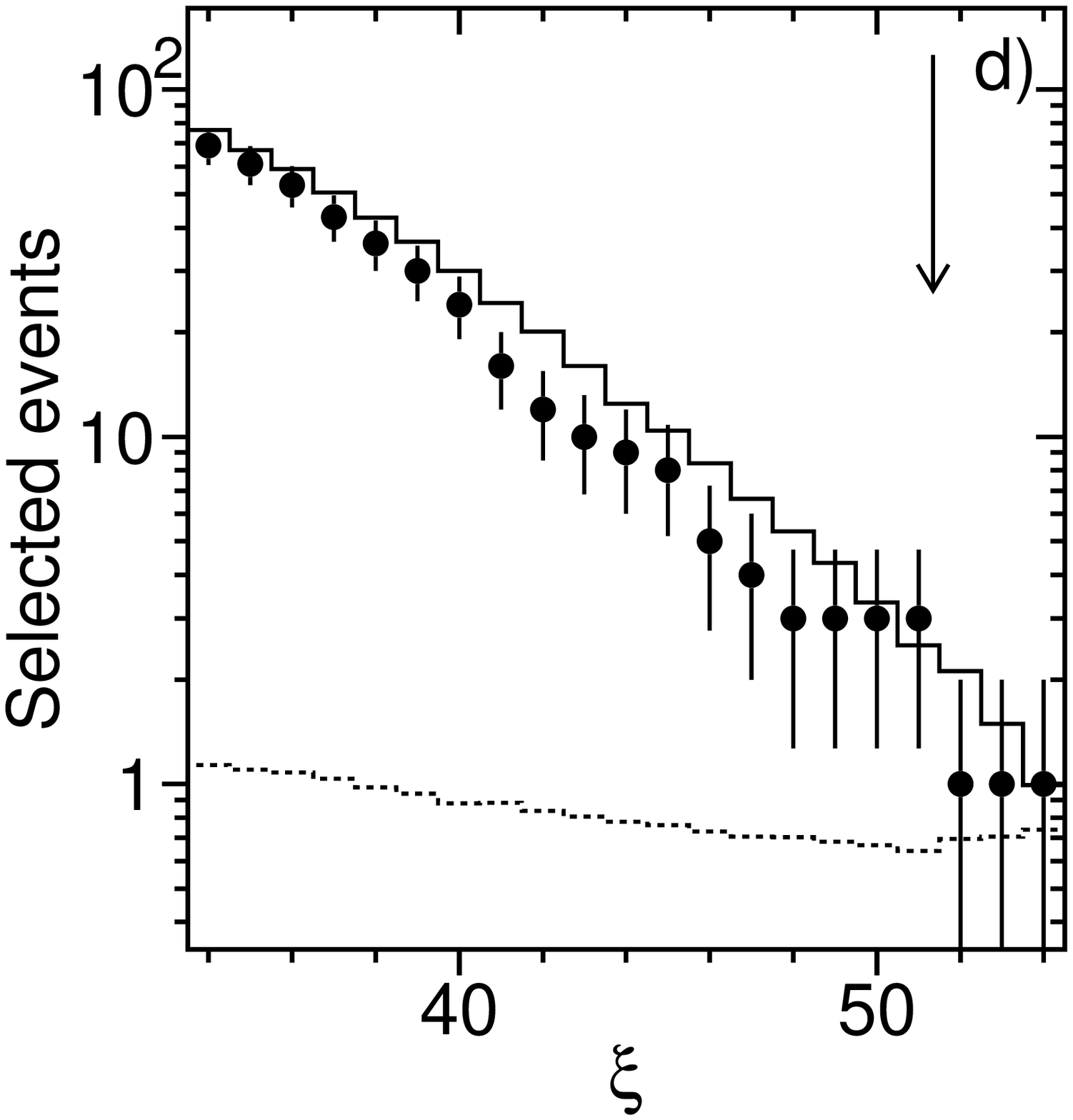,width=8.0cm} 
\end{tabular}
 \caption{
Number of events selected in data (points),
 in Monte Carlo simulation of standard processes (solid line)
and signal sensitivity (dashed line)
as a function of selection cuts with increasing background rejection
power. The vertical arrows show the $\xi$ value corresponding to
the optimised cuts.
The distributions are shown for the 
chargino lepton-jets low \dm\ a), the 
chargino lepton-jets medium \dm\ b), the
neutralino jet-jet very low \dm\ c) and the 
neutralino jet-jet high  \dm\ d)
selections, respectively
         }
 \label{fig:xi_chaneu}
\end{figure}

Each selection is parametrised as a function of a single parameter, $\xi$,
in the following manner: 
given a lower edge, $X_{loose}^i$, and an upper 
edge, $X_{tight}^i$,
for the cut on the variable $i$, the parameter $\xi$ 
is equal to 0 when this cut is at the lower edge
(many background events satisfy the selection)
and 100 when it is at the upper edge
(no or few background events pass the selection). All cuts  
($i = 1, . . ., N$) are related to the parameter $\xi$ as follows:
$$X_{cut}^i=
X_{loose}^i +
(X_{tight}^i- X_{loose}^i)
\times \frac{\xi}{100}.
$$
The parameter $\xi$ is scanned around the optimal value ($\xi$=50)
to check the agreement between data and Monte Carlo at different background 
rejection stages.
As illustrated in Figure~\ref{fig:xi_chaneu} 
for the lepton and hadrons final state in chargino decays and 
the pure hadronic final state in the neutralino decays, the data and
Monte Carlo simulations are 
in good agreement for all the \dm\ selections.
The vertical arrows show the $\xi$ value corresponding to
the optimised cuts.

For intermediate \dm\, values different from those chosen for
optimisation we choose the combination of selections 
providing the highest sensitivity \cite{susy_96}.
In this combination procedure, we take into account the overlap
among the selections within the 
data and Monte Carlo samples.

Typical selection efficiencies, as well as the number of background events
expected for a chargino mass of 94 \gev{} for the purely leptonic final state
(LL) or
for the $\Wstar\neutralino{1}$ decay mode, are displayed in
Table \ref{tab4}. In 
the latter case, a maximum efficiency of 47\% is reached for a background
contamination of 7.5 events for $\dm = 30 \gev{}$.
In the low \dm\ region the efficiency decreases due to
the large contamination of two-photon interactions and due to the
lower trigger acceptance.
For large \dm\ it decreases because of the WW background.

The selection efficiencies, as well as the number of background events
expected for a sum of neutralino masses 
$M_{\neutralino{1}} + M_{\neutralino{2}} = 188 \gev{}$ 
for the pure leptonic decays and for the $\Zstar\neutralino{1}$ 
decay mode are displayed in Table~\ref{tab5b}.
Compared to the chargino selection, the efficiencies are
lower due to the invisible decays of the \Zstar.

Systematic errors on the signal efficiencies are evaluated as in
Reference \citen{reflep1.5}, and they are typically 5\% relative, 
dominated by Monte Carlo statistics. These errors
are taken into account following the procedure explained 
in Reference \citen{cal_limit}.

\section{Model independent upper limits on production cross sections}
No excess of events is observed
and we set upper limits on the 
chargino and neutralino production cross sections 
in the framework of the MSSM.
Exclusion limits at 95\% C.L. are derived taking into account background
contributions.

To derive the upper limits on production cross sections 
and for interpretations in the MSSM we combine the 
$\sqrt{s} = 189 \gev{}$ data sample
with those collected by L3 at lower centre-of-mass energies 
\cite{reflep1.5,susy_96,susy_97}.

\begin{figure}[hbtp]
\begin{center}
  \mbox{\epsfxsize=10.0cm \epsffile{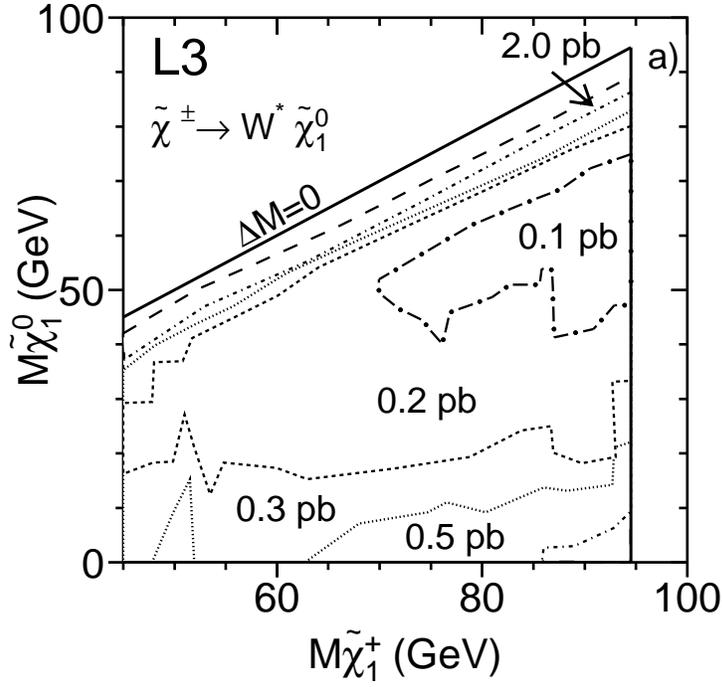}} \\ 
  \mbox{\epsfxsize=10.0cm \epsffile{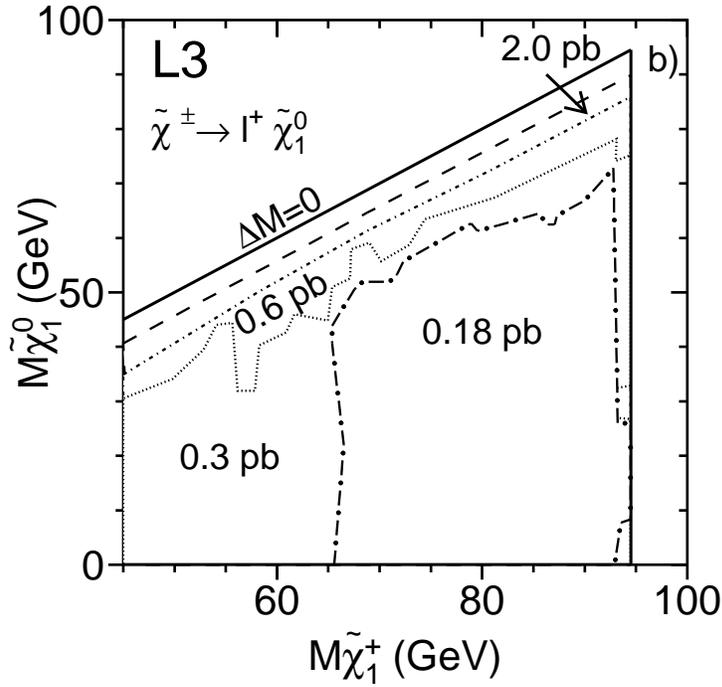}}
  \caption{Upper limits on the $\ee \rightarrow \chargino{+}\chargino{-}$
 production cross section
up to $\sqrt{s} = 189 \gev{}$ in 
the $M_{\neutralino{1}} - M_{\chargino{\pm}}$ plane.
           Exclusion limits are obtained assuming
  standard W branching
ratios in the chargino decay a) or purely leptonic W 
           decays b), 
           $\chargino{\pm} \rightarrow  \neutralino{1} \ell^\pm \nu $
           ($\ell=$e, $\mu$, $\tau$).
          }
  \label{fig:xsection_chargino}
 \end{center}
\end{figure}

The contours of upper limits on the production cross sections for the process
$\ee \rightarrow \chargino{\pm}\chargino{\mp}$ are shown in 
Figure~\ref{fig:xsection_chargino} assuming 
$\chargino{\pm} \rightarrow \Wstar\neutralino{1}$ for the chargino decay 
with standard W branching fractions, 
and for purely leptonic W decays.
In most of the kinematically accessible region, cross sections larger than 
0.2~pb are excluded for both scenarios.

\begin{figure}[hbtp]
 \begin{center}
  \mbox{\epsfxsize=10.0cm \epsffile{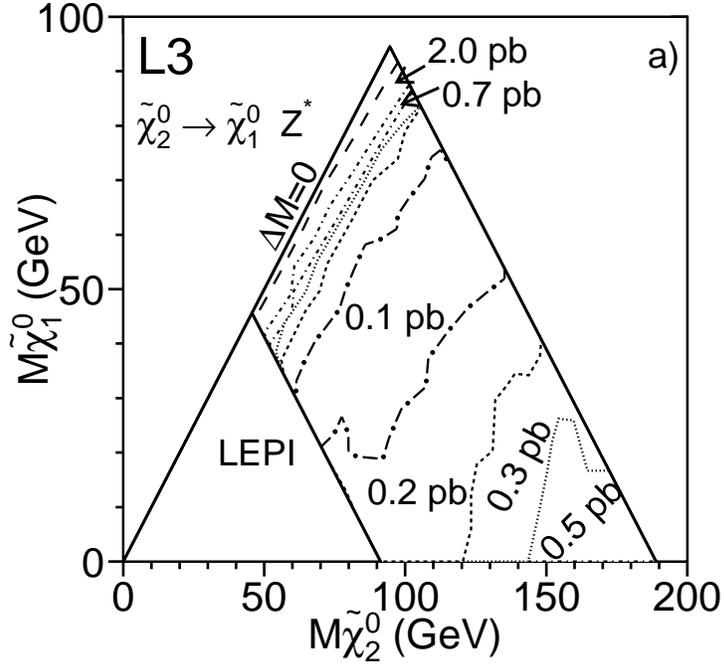}} \\
  \mbox{\epsfxsize=10.0cm \epsffile{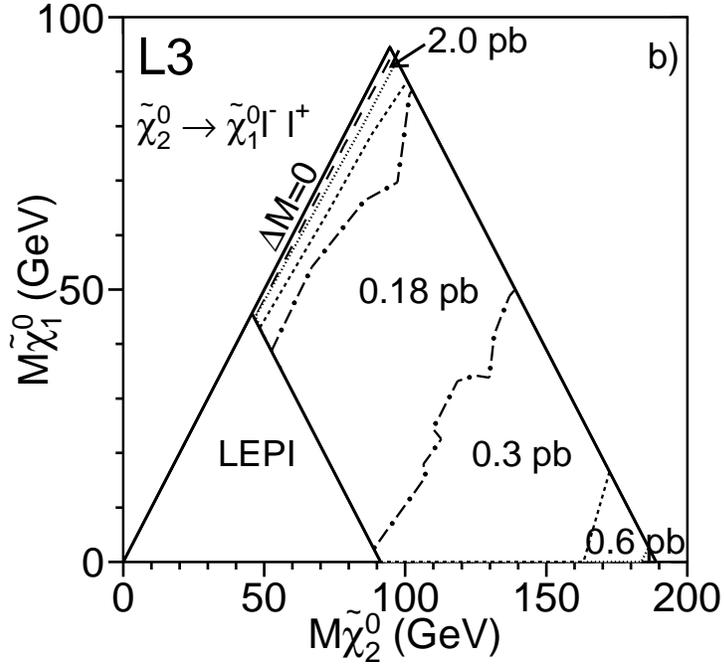}}
  \caption{Upper limits on the $\ee \rightarrow \neutralino{1}\neutralino{2}$
 production cross section
up to $\sqrt{s} = 189 \gev{}$ 
           in the $M_{\neutralino{1}} - M_{\neutralino{2}}$ plane.
           Exclusion limits are obtained 
assuming standard Z branching
 ratios in the next-to-lightest neutralino decay 
$\neutralino{2} \rightarrow \Zstar \neutralino{1}$ a)
or assuming purely leptonic Z 
           decays b),   
            $ \neutralino{2}\rightarrow  \neutralino{1} \ell^+ \ell^- $
           ($\ell=$e, $\mu$, $\tau$).
          }
  \label{fig:xsection_neutralino}
 \end{center}
\end{figure}

Similarly, cross section limits for associated neutralino production 
$\ee \rightarrow \neutralino{1}\neutralino{2}$ are derived as shown in
Figure~\ref{fig:xsection_neutralino}
assuming  $\neutralino{2} \rightarrow \Zstar\neutralino{1}$, 
with standard Z branching fractions 
and for purely leptonic Z decays.
In most of the kinematically accessible region, cross sections larger than 
0.3~pb are excluded for both scenarios.

\section{Interpretation in the MSSM}

In the MSSM, with Grand Unification assumptions~\cite{MSSM_GUT},
the masses and couplings of the SUSY
particles as well as their production cross sections,
are entirely described~\cite{mssm} once five parameters are fixed:
$\tan\beta$, the ratio of the vacuum expectation values of the two Higgs 
doublets, $M \equiv M_2$, the gaugino mass parameter,
$\mu$, the higgsino mixing parameter,
$m_0$, the common mass for scalar fermions at the GUT scale, and $A$, 
the trilinear coupling in the Higgs sector.
The following MSSM parameter space is investigated: 
$$
   \begin{array}{rclcrcl}
         0.7 \leq& \tan\beta & \leq 60 ,     &&
0 \leq &M_2 &\leq 2000 \gev{} ,\\
 -2000 \gev{} \leq& \mu       & \leq 2000 \gev{} ,&& 
0 \leq& m_0         &\leq  500 \gev{} .
  \end{array}
$$

To derive the absolute limits on the masses 
of the lightest neutralino and of
the lightest chargino,
a scan in the MSSM parameter space is performed 
in steps of $0.2 \gev{}$ for $M_2$,
$1.0 \gev{}$ for $\mu$ and $0.5 \gev{}$ for $m_0$. 

Mass eigenstates of scalar quarks and leptons are in general a 
mixture of the weak eigenstates $\susy{f}_R$ and $\susy{f}_L$.
The mixing between these two states
is proportional to the mass of the partner fermion.
Hence the mixing can be sizable only for particles of the
third generation. The mixing is governed by the parameters $A$, 
$\mu$ and $\tan\beta$. Besides $\mu$ and $\tan\beta$, also
a scan on $A$ is performed to check the validity of the following results.

All the limits on the cross sections previously shown, combined with
the results obtained at lower centre-of-mass energies and with the
results of scalar lepton searches obtained at $\sqrt{s} = 189 \gev{}$ 
\cite{slep99ref}, can be 
translated into exclusion regions in the MSSM parameter space.
To derive limits in the MSSM, 
we optimise the global selection for any different
point in the parameter space. This is obtained, choosing every time the
combination of selections providing the highest sensitivity, given 
the production cross sections and the decay branching fractions which are 
calculated with the generator {\tt SUSYGEN}. When the mixing in the
scalar tau sector is considered, masses and decay branching fractions
are calculated with the generator {\tt ISAJET} \cite{isajet}.

\subsection{Limits on chargino and neutralino masses} 

In the MSSM, while the cross sections and decay branching fractions
of the
charginos and neutralinos depend on the masses of the scalar leptons, 
their masses depend only on 
$M_2$, $\mu$ and $\tan\beta$. 
The exclusions in the
high $m_0$ range are derived from chargino and
neutralino searches, while for low $m_0$ the searches for scalar
leptons \cite{slep99ref}, and for photons and
missing energy final states \cite{papho97}, also contribute.
We also take into account all chargino and neutralino cascade decays:
\begin{itemize}
\item
 $\chargino{\pm} \rightarrow \neutralino{2}\, \Wstar$: 
 we observe a slight decrease of the efficiency relative to
 $\chargino{\pm} \rightarrow \neutralino{1}\, \Wstar$
 depending on the masses of \neutralino{2}, \chargino{\pm} and 
 \neutralino{1}.
 The lowest efficiency is then used for cascade decays.
\item
 $\neutralino{3,4} \rightarrow \neutralino{2}\, \Zstar$:
 the efficiency is found to be larger than the efficiency obtained for the
 $\neutralino{3,4} \rightarrow \neutralino{1}\, \Zstar$ channel, 
 especially in the high \dm\ region. 
 The efficiencies obtained in the latter
 channel are used. 
\item 

$\neutralino{3,4} \rightarrow \sneutrino \nu$: when the $ \sneutrino$
becomes detectable through its cascade decays into $\neutralino{2}$ or 
  $\chargino{\pm}$. This is especially relevant in the mixed region
($\mu\sim -M_2$) for
the low $\tan\beta$ values.
\end{itemize}

\begin{figure}[hbtp]
 \begin{center} \vspace*{-1.5cm}
 \hspace*{-1.5cm}\mbox{\epsfxsize=11.50cm \epsffile{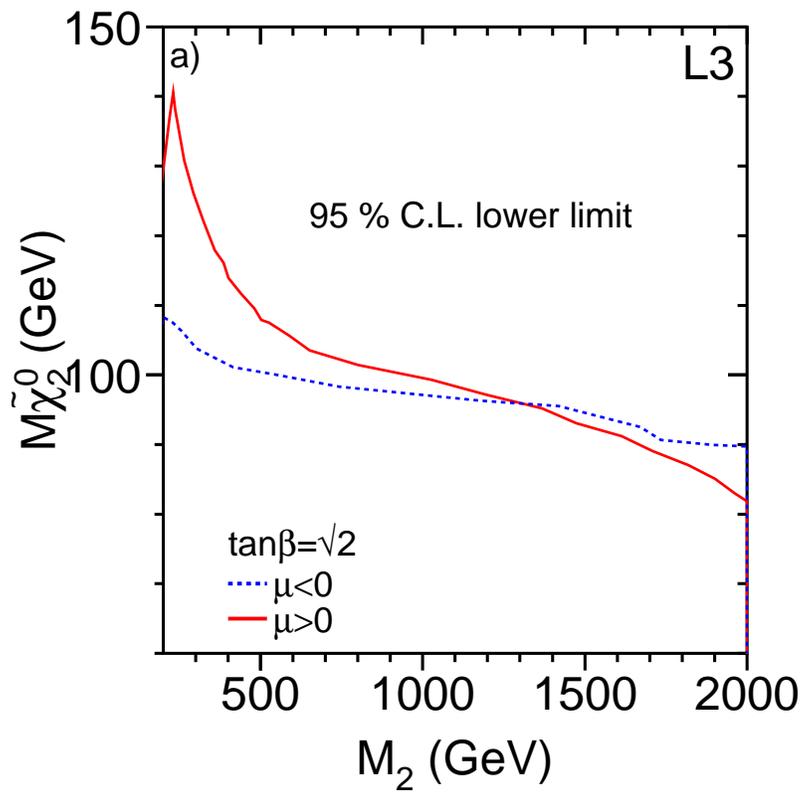}} \\
  \hspace*{-1.5cm}\mbox{\epsfxsize=11.50cm \epsffile{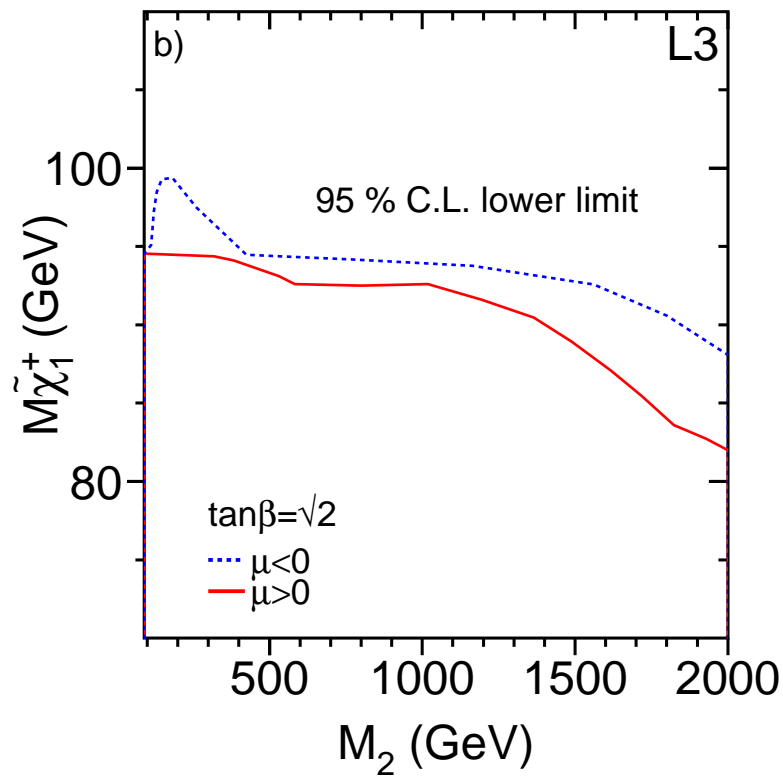}}
\vspace{-.5cm}
 \caption {Lower mass limits as a function of $M_2$ for
           the next-to-lightest neutralino a)
and the lightest chargino b).
 The limits are shown for $\tan\beta = \sqrt{2}$ and for $\mu>0$ and $\mu<0$. 
          } 
  \label{fig:mass_m2}
 \end{center}
\end{figure}

Depending on the neutralino-chargino field content, 
one distinguishes the following 
cases for the determination of lower limits on the neutralino and chargino 
masses:
\begin{itemize}
\item   
 Higgsino-like  \neutralino{2} and \chargino{\pm}  ($M_2 \gg |\mu|$):
 in this case, the production cross sections do not depend on the scalar lepton
 masses, \dm\ is low and decreases with increasing $M_2$.
Consequently, the limits on the masses of the next-to-lightest neutralino
 and the lightest chargino decrease with $M_2$ as depicted in 
 Figure~\ref{fig:mass_m2}.
 For $\tan\beta = \sqrt{2}$ and $M_2$ less than $500 \gev{}$, 
$M_{\neutralino{2}} \leq 101 \GeV{}$ and
 $M_{\chargino{\pm}} \leq 93 \GeV{}$
 are excluded.
\item
 Gaugino-like $\chargino{\pm}$  ($|\mu| \gg M_2$):
 the chargino cross section depends strongly on the scalar neutrino mass. 
 For $50 \gev{} \leq M_{\sneutrino} \leq 80 \gev{}$ 
the cross section is reduced by one order of magnitude compared to what is
 expected for $M_{\sneutrino} \geq 500 \gev{}$.
  When the two body decay 
 $\chargino{\pm} \rightarrow \ell^\pm \, \sneutrino$ 
 is dominant, the relevant \dm\ becomes 
 $\dm = M_{\chargino{\pm}} - M_{\sneutrino}$.
If the $\sneutrino$ is mass degenerate with the \chargino{\pm} the 
acceptance is substantially reduced. However, when this occurs scalar 
leptons are light and the experimental sensitivity is recovered with these
channels.

\end{itemize}

\begin{figure}[hbtp]
 \begin{center}

  \hspace*{-1.5cm}\mbox{\epsfxsize=18cm \epsffile{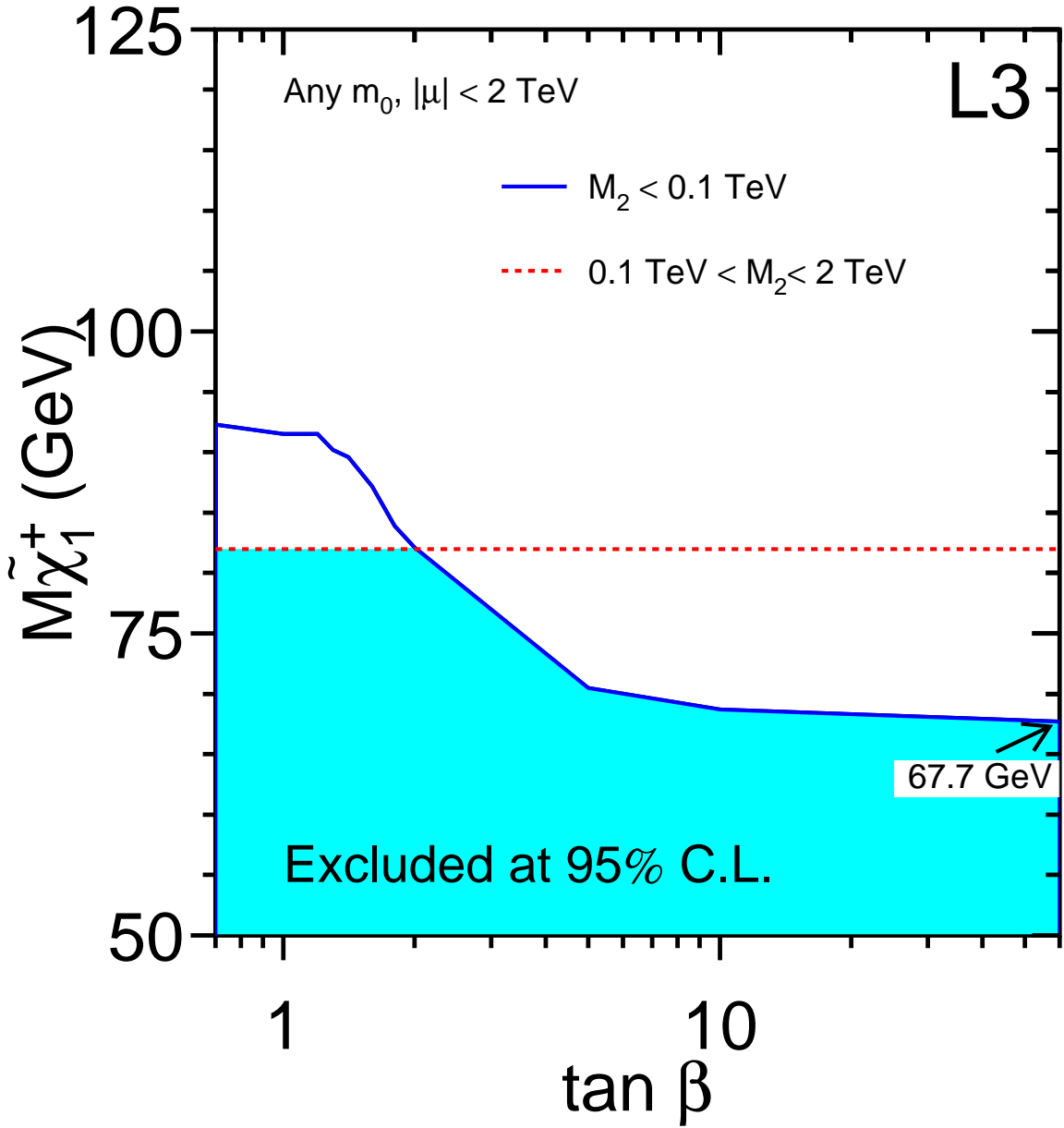}}
  \caption{
Lower limit on $M_{\chargino{\pm}}$ as a function of
           $\tan\beta$ and for any value of $m_0$.
The solid line (gaugino region) shows the lower limit obtained for
light scalar neutrinos (also small $M_2$ values), which corresponds to the 
absolute lower limit for large $\tan\beta$ values. The 
dashed line (higgsino region) shows the lower limit obtained for very small $\dm$ values. This line corresponds to the absolute lower limit for 
small $\tan\beta$ values.
  \label{fig:mass_cha}}
 \end{center}
\end{figure}

The mass limit of the lightest
chargino is shown in Figure \ref{fig:mass_cha} as a function of 
$\tan\beta$ for all the different chargino field contents. 
At large $\tan\beta$ values,
the lower mass limit of the lightest chargino is obtained
when the lightest chargino and the $\sneutrino$ are mass degenerate
(gaugino region). At 
low $\tan\beta$ values the lower mass limit on the lightest 
chargino is obtained when the lightest chargino and 
the $\neutralino{1}$ (LSP) are mass
degenerate (higgsino region). Finally, for $M_2<2$ \TeV, 
for $\tan\beta \leq 60$ and for any $m_0$ values, 
the lower mass limit of the lightest chargino is:
$$   M_{\chargino{\pm}} \geq 67.7 \gev{} .$$

The scalar tau can be much lighter than
the scalar electron and muon. This mass splitting 
occurs in particular for large $\tan\beta$ and $A$ values.
When this happens, chargino and next-to-lightest neutralino decays 
are affected. Therefore, detection efficiencies are estimated for chargino and
next-to-lightest neutralino decays with 100\% branching ratio
into $\stau{}_{1}\nu$ and $\stau{\pm}_{1}\tau^{\mp}$, respectively.
In particular, when the 
$\stau{}_{1}$ and the LSP are mass degenerate the efficiencies 
decrease substantially.
However, the experimental sensitivity can be partially recovered
taking into account also the process
$\epem\ra \susy{\nu}\susy{\nu}$, where 
the $\susy{\nu}$ is visible through its cascade decays.
In particular, the limit on the chargino mass 
holds for any value of the mixing if $\tan\beta < 20$.
For higher $\tan\beta$ values this limit can be decreased
at most by 10 \gev. 

\begin{figure}
 \begin{center}
\vspace{-1.5cm}
  \hspace*{-2.0cm}\mbox{\epsfxsize=11.5cm \epsffile{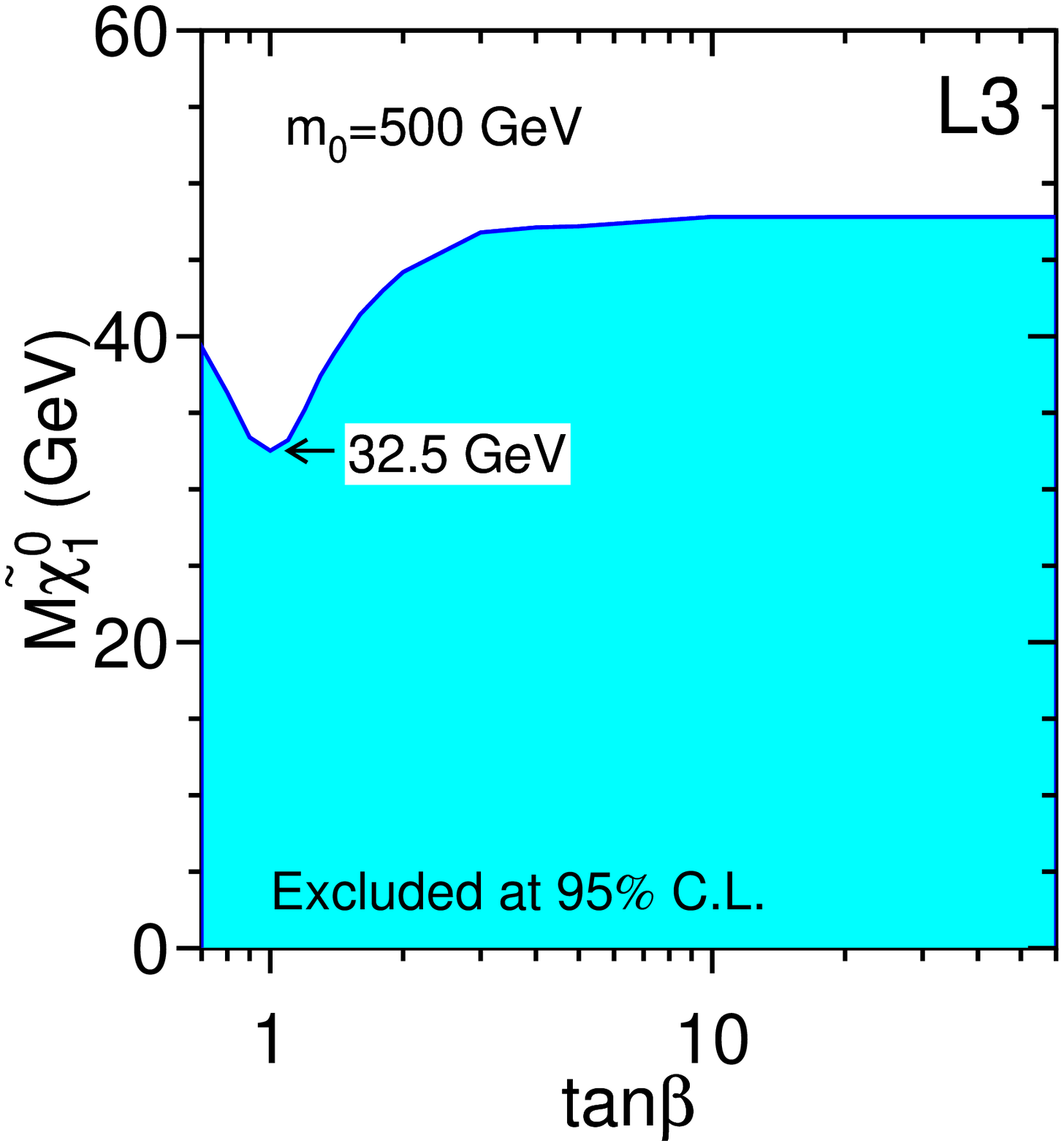}}
\vspace{-0.5cm}
 \caption{ Lower limit on the lightest neutralino mass, $M_{\neutralino{1}}$,
           as a function of $\tan\beta$ for  
           $m_0=500 \gev{}$, when combining all chargino and neutralino searches.}
 \label{fig:neutralino_mass_a}
\vspace{-.5cm}
  \hspace*{-1.5cm}\mbox{\epsfxsize=11.5cm \epsffile{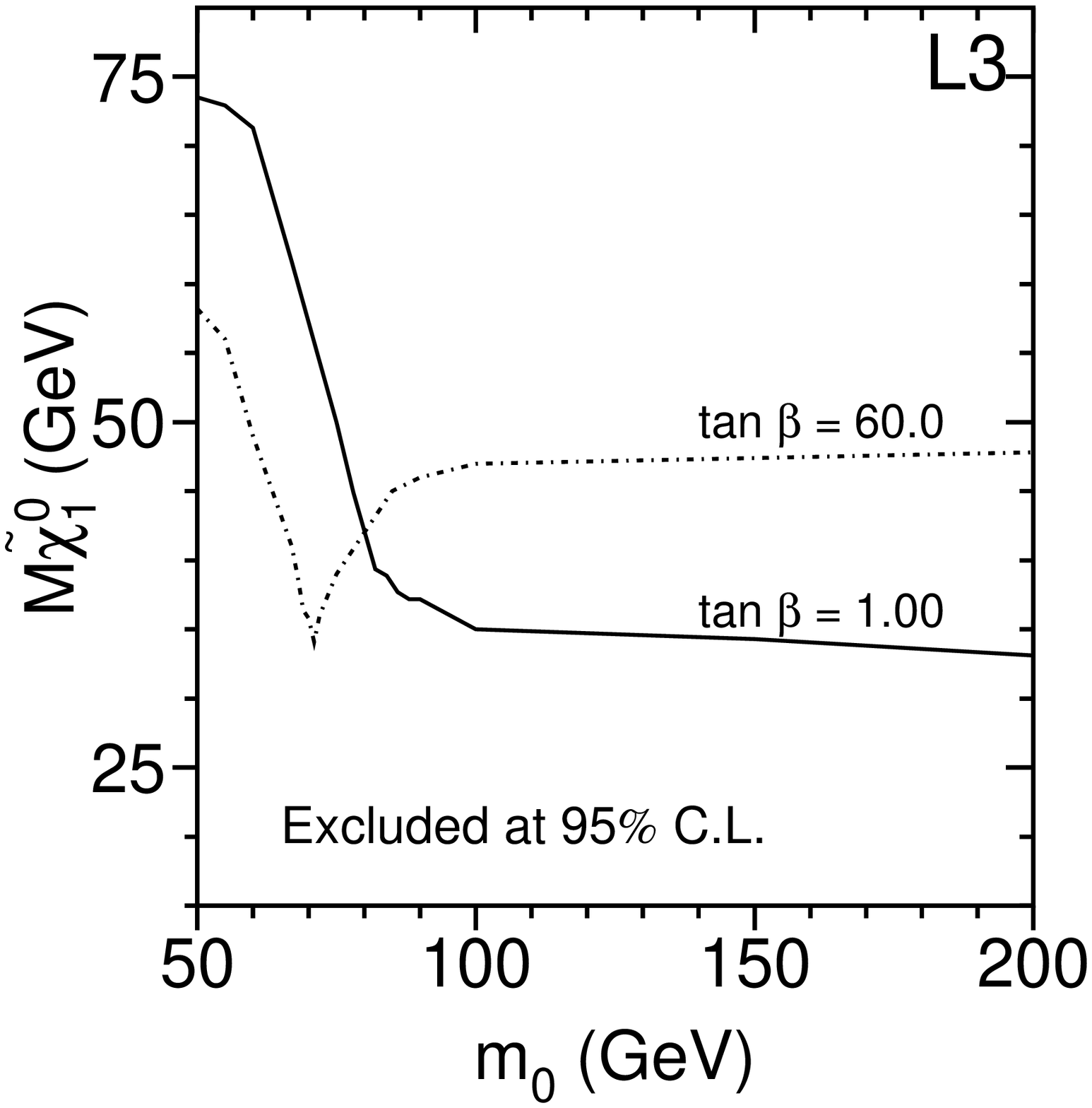}}
\vspace{0.0cm}
  \caption{Lower limit on the lightest neutralino mass, $M_{\neutralino{1}}$,
            as a function of
            $m_0$ for two values of $\tan\beta$. Scalar lepton searches
contribute in the low $m_0$ region. Chargino searches contribute mainly
in the high $m_0$ region. For the low $\tan\beta$ values, the neutralino
searches give additional contribution in the intermediate $m_0$ region.
          }
  \label{fig:neutralino_mass_b}
 \end{center}
\end{figure}

Indirect limits on the mass of the lightest neutralino
are also derived as a function of $m_0$ and as a function of tan$\beta$.
In the
low  $m_0$ region ($\leq 65 \gev{}$) the mass limit on the LSP
comes mainly from the scalar lepton searches.
For large $m_0$ values ($\geq 200 \gev{}$), only  
the chargino and neutralino searches contribute. At 
low $\tan\beta$ the processes
$\epem\ \rightarrow \neutralino{2} \neutralino{3,4}$
contribute significantly and they are taken into account.
The lower mass limit is found at $\tan\beta = 1$, $\mu = -70 \gev{}$
and $m_0 = 500 \gev{}$, as shown in Figure~\ref{fig:neutralino_mass_a}.
For these values of the parameters, 
the chargino mass is at the kinematic limit and 
the mass difference between the chargino 
and the LSP is maximal.

For intermediate $m_0$ values ($65 \gev{} \leq m_0 \leq 95 \gev{}$)
the production cross section for charginos is minimal
and the $\sneutrino$ is light enough to allow
the following decay modes:
$\neutralino{2,3,4} \rightarrow \sneutrino\nu$ and
$\chargino{\pm} \rightarrow \sneutrino \ell^{\pm}$. 
This is the region where the exclusion is due to the interplay of many
different searches. The limit on the lightest neutralino as a function
of $m_0$, and for two extreme values of $\tan\beta$, 
is shown in Figure~\ref{fig:neutralino_mass_b}. 
For low $\tan\beta$ values
($\le \sqrt{2}$), the minimum is found for
$\mu \sim -70 \gev{}$ and large $m_0$ values. 
Better limits are obtained  
for intermediate $m_0$ values, where 
the neutralino production cross sections are large and the 
two body decays of the $\neutralino{3,4}$ into $\sneutrino \nu $   
are visible through the cascade decays of the
$\sneutrino$. 
For larger $\tan\beta$ values, the minimum is found in the gaugino 
region ($-2000 \gev{}  < \mu < -200 \gev{}$) and for 
$70 \gev{} \leq m_0 \leq 80 \gev{}$. In this region of the parameter space,   
the $\sneutrino$ and the chargino are 
mass degenerate, the heavier neutralinos  
decay invisibly and the experimental sensitivity 
is entirely due to the scalar lepton searches.

\begin{figure}
 \begin{center}
\vspace{-1.cm}
  \hspace*{-1.5cm}\mbox{\epsfxsize=11.5cm \epsffile{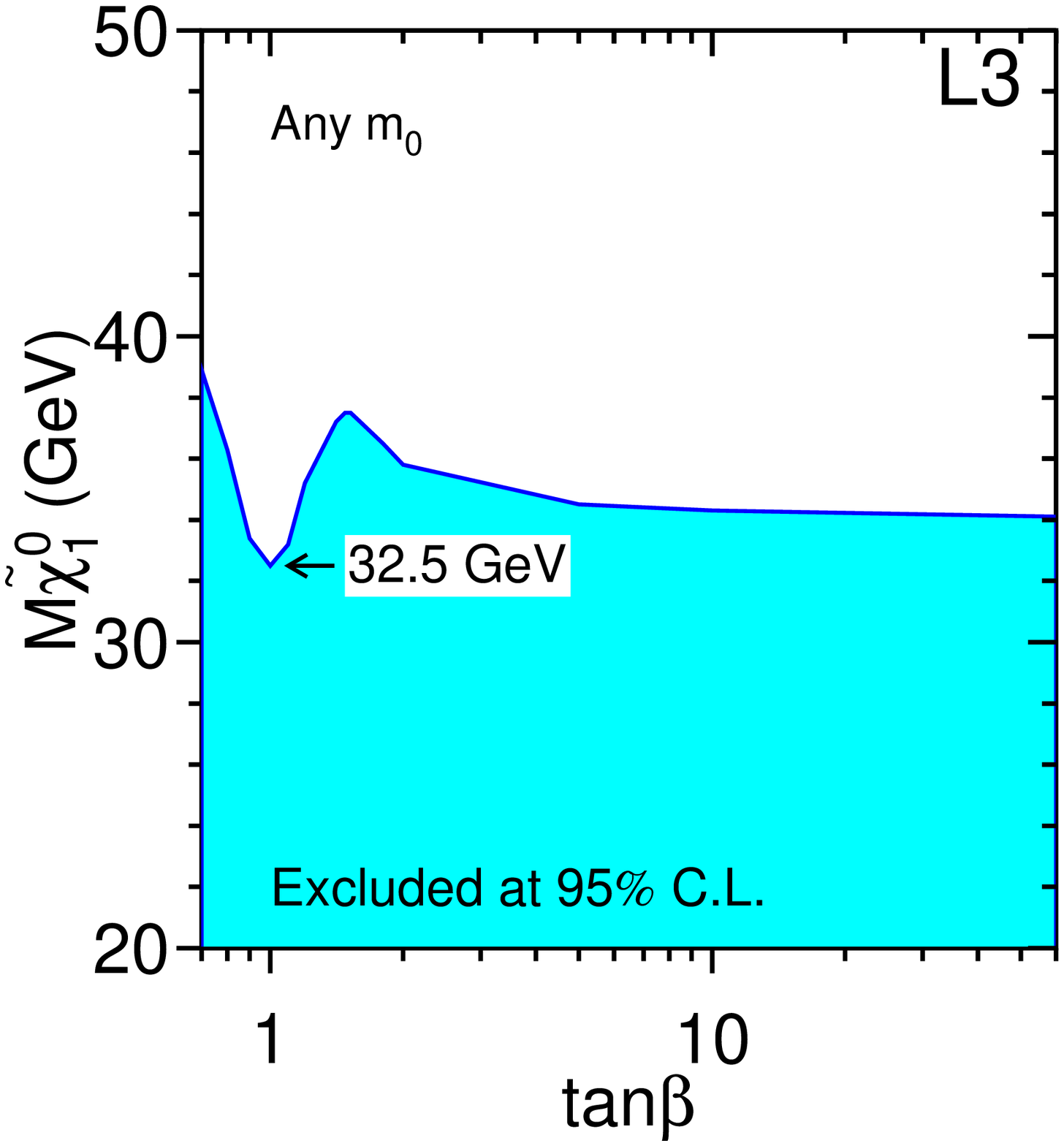}}
\vspace{-.5cm}
  \caption{Lower limit on $M_{\neutralino{1}}$
            as a function of
  $\tan\beta$ and for any value of $m_0$,  when combining
the chargino, neutralino and scalar lepton searches.
          }
  \label{fig:neutralino_any_a}
\vspace{-.5cm}
  \hspace*{-1.5cm}\mbox{\epsfxsize=11.5cm \epsffile{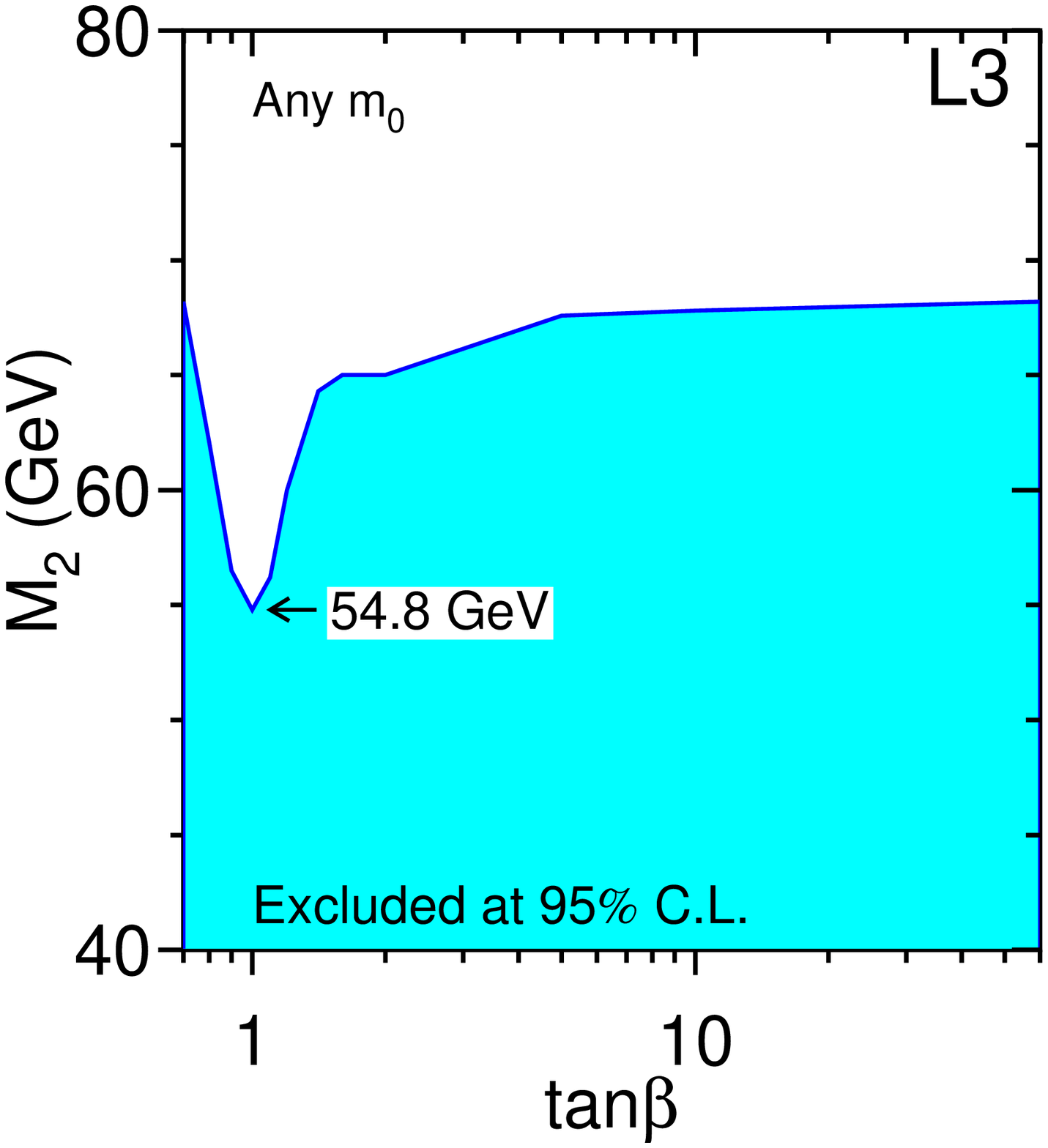}}
  \caption{Lower limit on $M_2$
            as a function of
           $\tan\beta$ and for any value of $m_0$,  when combining
the chargino neutralino and scalar lepton searches.          }
  \label{fig:neutralino_any_b}
 \end{center}
\end{figure}

Finally in Figure~\ref{fig:neutralino_any_a}, the mass limit on the
lightest neutralino as a function of $\tan\beta$ for any $m_0$ value
is shown. For $\tan\beta \geq 0.7$, the 
lower mass limit of the lightest neutralino is 
$$
   M_{\neutralino{1}} \geq 32.5 \gev{} .
$$

The mass limit on the
lightest neutralino is very little affected
by the mixing in the scalar tau sector.
The limit holds for any value of the mixing if $\tan\beta < 20$
and it can be reduced at most by 1.5 \gev{} for higher $\tan\beta$ values.
Nevertheless, the absolute mass limit for the lightest neutralino
does not change since the lowest value is still found at $\tan\beta =1$.


The mass limit on the lightest neutralino is also translated in
an absolute limit on $M_2$. This is shown as a function 
of $\tan\beta$ for any $m_0$ and $\mu$ as depicted 
in Figure ~\ref{fig:neutralino_any_b}. 
Values of $M_2$ lower than $54.8 \gev{}$ are now excluded at 95$\%$ C.L.


\section*{Acknowledgments}

We  express our gratitude to the CERN accelerator divisions for the
excellent performance of the LEP machine. We also acknowledge
and appreciate the effort of the engineers, technicians and support staff 
who have participated in the construction and maintenance of this experiment.

\newpage


\bibliographystyle{l3stylem}

%
%
\newpage
\typeout{   }     
\typeout{Using author list for paper 186 -?}
\typeout{$Modified: Fri Sep 10 08:43:14 1999 by clare $}
\typeout{!!!!  This should only be used with document option a4p!!!!}
\typeout{   }
%
%
%
%
%
%

\newcount\tutecount  \tutecount=0
\def\tutenum#1{\global\advance\tutecount by 1 \xdef#1{\the\tutecount}}
\def\tute#1{$^{#1}$}
\tutenum\aachen            
\tutenum\nikhef            
\tutenum\mich              
\tutenum\lapp              
\tutenum\basel             
\tutenum\lsu               
\tutenum\beijing           
\tutenum\berlin            
\tutenum\bologna           
\tutenum\tata              
\tutenum\ne                
\tutenum\bucharest         
\tutenum\budapest          
\tutenum\mit               
\tutenum\debrecen          
\tutenum\florence          
\tutenum\cern              
\tutenum\wl                
\tutenum\geneva            
\tutenum\hefei             
\tutenum\seft              
\tutenum\lausanne          
\tutenum\lecce             
\tutenum\lyon              
\tutenum\madrid            
\tutenum\milan             
\tutenum\moscow            
\tutenum\naples            
\tutenum\cyprus            
\tutenum\nymegen           
\tutenum\caltech           
\tutenum\perugia           
\tutenum\cmu               
\tutenum\prince            
\tutenum\rome              
\tutenum\peters            
\tutenum\salerno           
\tutenum\ucsd              
\tutenum\santiago          
\tutenum\sofia             
\tutenum\korea             
\tutenum\alabama           
\tutenum\utrecht           
\tutenum\purdue            
\tutenum\psinst            
\tutenum\zeuthen           
\tutenum\eth               
\tutenum\hamburg           
\tutenum\taiwan            
\tutenum\tsinghua          
{
\parskip=0pt
\noindent
{\bf The L3 Collaboration:}
\ifx\selectfont\undefined
 \baselineskip=10.8pt
 \baselineskip\baselinestretch\baselineskip
 \normalbaselineskip\baselineskip
 \ixpt
\else
 \fontsize{9}{10.8pt}\selectfont
\fi
\medskip
\tolerance=10000
\hbadness=5000
\raggedright
\hsize=162truemm\hoffset=0mm
\def\r{\rlap,}
\noindent

M.Acciarri\r\tute\milan\
P.Achard\r\tute\geneva\ 
O.Adriani\r\tute{\florence}\ 
M.Aguilar-Benitez\r\tute\madrid\ 
J.Alcaraz\r\tute\madrid\ 
G.Alemanni\r\tute\lausanne\
J.Allaby\r\tute\cern\
A.Aloisio\r\tute\naples\ 
M.G.Alviggi\r\tute\naples\
G.Ambrosi\r\tute\geneva\
% ...
\newpage
\begin{list}{A}{\itemsep=0pt plus 0pt minus 0pt\parsep=0pt plus 0pt minus 0pt
                \topsep=0pt plus 0pt minus 0pt}
\item[\aachen]
 I. Physikalisches Institut, RWTH, D-52056 Aachen, FRG$^{\S}$\\
 III. Physikalisches Institut, RWTH, D-52056 Aachen, FRG$^{\S}$
\item[\nikhef] National Institute for High Energy Physics, NIKHEF, 
     and University of Amsterdam, NL-1009 DB Amsterdam, The Netherlands
\item[\mich] University of Michigan, Ann Arbor, MI 48109, USA
\item[\lapp] Laboratoire d'Annecy-le-Vieux de Physique des Particules, 
     LAPP,IN2P3-CNRS, BP 110, F-74941 Annecy-le-Vieux CEDEX, France
\item[\basel] Institute of Physics, University of Basel, CH-4056 Basel,
     Switzerland
\item[\lsu] Louisiana State University, Baton Rouge, LA 70803, USA
\item[\beijing] Institute of High Energy Physics, IHEP, 
  100039 Beijing, China$^{\triangle}$ 
\item[\berlin] Humboldt University, D-10099 Berlin, FRG$^{\S}$
\item[\bologna] University of Bologna and INFN-Sezione di Bologna, 
     I-40126 Bologna, Italy
\item[\tata] Tata Institute of Fundamental Research, Bombay 400 005, India
\item[\ne] Northeastern University, Boston, MA 02115, USA
\item[\bucharest] Institute of Atomic Physics and University of Bucharest,
     R-76900 Bucharest, Romania
\item[\budapest] Central Research Institute for Physics of the 
     Hungarian Academy of Sciences, H-1525 Budapest 114, Hungary$^{\ddag}$
\item[\mit] Massachusetts Institute of Technology, Cambridge, MA 02139, USA
\item[\debrecen] Lajos Kossuth University-ATOMKI, H-4010 Debrecen, Hungary$^\P$
\item[\florence] INFN Sezione di Firenze and University of Florence, 
     I-50125 Florence, Italy
\item[\cern] European Laboratory for Particle Physics, CERN, 
     CH-1211 Geneva 23, Switzerland
\item[\wl] World Laboratory, FBLJA  Project, CH-1211 Geneva 23, Switzerland
\item[\geneva] University of Geneva, CH-1211 Geneva 4, Switzerland
\item[\hefei] Chinese University of Science and Technology, USTC,
      Hefei, Anhui 230 029, China$^{\triangle}$
\item[\seft] SEFT, Research Institute for High Energy Physics, P.O. Box 9,
      SF-00014 Helsinki, Finland
\item[\lausanne] University of Lausanne, CH-1015 Lausanne, Switzerland
\item[\lecce] INFN-Sezione di Lecce and Universit\'a Degli Studi di Lecce,
     I-73100 Lecce, Italy
\item[\lyon] Institut de Physique Nucl\'eaire de Lyon, 
     IN2P3-CNRS,Universit\'e Claude Bernard, 
     F-69622 Villeurbanne, France
\item[\madrid] Centro de Investigaciones Energ{\'e}ticas, 
     Medioambientales y Tecnolog{\'\i}cas, CIEMAT, E-28040 Madrid,
     Spain${\flat}$ 
\item[\milan] INFN-Sezione di Milano, I-20133 Milan, Italy
\item[\moscow] Institute of Theoretical and Experimental Physics, ITEP, 
     Moscow, Russia
\item[\naples] INFN-Sezione di Napoli and University of Naples, 
     I-80125 Naples, Italy
\item[\cyprus] Department of Natural Sciences, University of Cyprus,
     Nicosia, Cyprus
\item[\nymegen] University of Nijmegen and NIKHEF, 
     NL-6525 ED Nijmegen, The Netherlands
\item[\caltech] California Institute of Technology, Pasadena, CA 91125, USA
\item[\perugia] INFN-Sezione di Perugia and Universit\'a Degli 
     Studi di Perugia, I-06100 Perugia, Italy   
\item[\cmu] Carnegie Mellon University, Pittsburgh, PA 15213, USA
\item[\prince] Princeton University, Princeton, NJ 08544, USA
\item[\rome] INFN-Sezione di Roma and University of Rome, ``La Sapienza",
     I-00185 Rome, Italy
\item[\peters] Nuclear Physics Institute, St. Petersburg, Russia
\item[\salerno] University and INFN, Salerno, I-84100 Salerno, Italy
\item[\ucsd] University of California, San Diego, CA 92093, USA
\item[\santiago] Dept. de Fisica de Particulas Elementales, Univ. de Santiago,
     E-15706 Santiago de Compostela, Spain
\item[\sofia] Bulgarian Academy of Sciences, Central Lab.~of 
     Mechatronics and Instrumentation, BU-1113 Sofia, Bulgaria
\item[\korea] Center for High Energy Physics, Adv.~Inst.~of Sciences
     and Technology, 305-701 Taejon,~Republic~of~{Korea}
\item[\alabama] University of Alabama, Tuscaloosa, AL 35486, USA
\item[\utrecht] Utrecht University and NIKHEF, NL-3584 CB Utrecht, 
     The Netherlands
\item[\purdue] Purdue University, West Lafayette, IN 47907, USA
\item[\psinst] Paul Scherrer Institut, PSI, CH-5232 Villigen, Switzerland
\item[\zeuthen] DESY, D-15738 Zeuthen, 
     FRG
\item[\eth] Eidgen\"ossische Technische Hochschule, ETH Z\"urich,
     CH-8093 Z\"urich, Switzerland
\item[\hamburg] University of Hamburg, D-22761 Hamburg, FRG
\item[\taiwan] National Central University, Chung-Li, Taiwan, China
\item[\tsinghua] Department of Physics, National Tsing Hua University,
      Taiwan, China
\item[\S]  Supported by the German Bundesministerium 
        f\"ur Bildung, Wissenschaft, Forschung und Technologie
\item[\ddag] Supported by the Hungarian OTKA fund under contract
numbers T019181, F023259 and T024011.
\item[\P] Also supported by the Hungarian OTKA fund under contract
  numbers T22238 and T026178.
\item[$\flat$] Supported also by the Comisi\'on Interministerial de Ciencia y 
        Tecnolog{\'\i}a.
\item[$\sharp$] Also supported by CONICET and Universidad Nacional de La Plata,
        CC 67, 1900 La Plata, Argentina.
\item[$\diamondsuit$] Also supported by Panjab University, Chandigarh-160014, 
        India.
\item[$\triangle$] Supported by the National Natural Science
  Foundation of China.
\item[\dag] Deceased.
\end{list}
}
\vfill






\newpage
%


\begin{table} [htbp]
\begin{center} 
\begin{tabular}{|c|c|c|c|c|} \hline  
\multicolumn{5}{|c|}{Electron/Muon plus hadrons selections  } \\  \hline
           & Very Low \dm & Low \dm & Medium \dm & Large \dm \\  \hline
No. of isolated leptons   $\ge$ &   1   &   1     & 1      &  1    \\  \hline
$N_{tk} - N_{lep}$       $\ge$ &  2    &  4      & 5      &  4    \\  \hline
$N_{cl}$                 $\ge$ &  6    &  10     & 10     &  10   \\  \hline
$\sin(\theta_{miss})$    $\ge$ &  0.74 &  0.38   & 0.23   &  0.28 \\  \hline
 $E^{\perp}_{25}$ (\gev{})  $\le$ &  0.52 &  --     & --     & 11.6  \\  \hline
 $p_{\perp}$ (\gev{})       $\ge$ &  3.24 &  5.62   & 8.65   & 9.84  \\  \hline
 $E_{lep} $  (\gev{})       $\ge$ &  1.51 & 2.59    & 6.17   & 25.9  \\  \hline
 $E_{lep} $  (\gev{})       $\le$ &  9.12 &  27.5   & 31.2   & 43.8  \\  \hline
 $E_{TTJL} $  (\gev{})       $\ge$ &  1.27 &  0.95  & 1.44   &  --   \\  \hline
 $M_{had} $  (\gev{})       $\le$ &  5.0   &  28.2  & 39.1   & 89.0  \\  \hline
 $M_{rec} $  (\gev{})      $\ge$ &  144   &  130    & 107    & 57.0  \\  \hline
 $E_{vis} $  (\gev{})      $\ge$ &  4.02   & 8.90   & 31.5   & 65.3  \\  \hline
 $E_{vis} $  (\gev{})     $\le$ &  11.0  &  59.1   & 93.6   & 118   \\  \hline
\end{tabular} 
\caption{Values of the cuts for the lepton plus hadrons selections; they
         are determined with    the optimisation procedure described in 
         Section~\protect\ref{sec:optimization}.\label{tab2}}
\end{center}
\end{table}
\begin{table}[!]
\vspace{-0.1 cm}
\begin{center}
\begin{tabular}{|c|c|c|c|c|} \hline  
\multicolumn{5}{|c|}{Chargino Hadronic selections  } \\  \hline
                  & Very Low \dm    & Low \dm & Medium \dm & Large \dm \\  \hline
$N_{cl}$             $\ge$ &  14    & 14      &  14    &  14      \\  \hline
$N_{tk}$             $\ge$ &  5     & 5       & 5      &  5       \\  \hline
 $p_{\perp} $ (\gev{})   $\ge$ &  3.72  & 10.0    & 11.5   & 11.4     \\  \hline
 $p_{\perp}/E_{vis}$  $\ge$ &  --    & 0.20    & 0.15   & 0.10     \\  \hline
$E_{vis}$ (\gev{})      $\le$ & 12.0   &  68.0   & 76.0   & 149    \\  \hline
 Acollinearity (rad)  $\le$ & 2.00   & --      & --     &  3.02  \\  \hline
 Acoplanarity (rad)   $\le$ & 2.18   & 2.89    &  2.92  &  3.11 \\  \hline
 sin($\theta_{miss}$) $\ge$ &  0.56  & 0.46    &  0.20  &  0.61  \\  \hline
$E^{\perp}_{25} $ (\gev{}) $\le$ &  0.21& 5.80    &  5.80  & 3.25     \\  \hline
 $E_{25}$ (\gev{})        $\le$ &  --  & --      &  --    & 2.53     \\  \hline
 $p_{\parallel}/E_{vis}$  $\le$ &  --  & 0.53    &  0.95  &  0.55  \\  \hline
 $E_{lep}^{max}$ (\gev{}) $\le$ &  9.12 &  27.5   & 31.2   & 43.8  \\  \hline
 $M_{vis}$ (\gev{})       $\ge$ &  2.85& 9.3    &  35.4  &  --      \\  \hline
 $M_{rec}$ (\gev{})       $\ge$ &   -- & 124     &   67.2 & --     \\  \hline
 $E_{vis}/\sqrt{s}$     $\ge$ &   -- & --      &   --   & 0.60     \\  \hline
 $E_{30^0}/E_{vis}$     $\le$ &  --  & 0.22    & 0.40   & 0.65      \\  \hline
 $E_{TTJ}/p_\perp$      $\ge$ &  0.24& --      &  0.24  & --       \\  \hline
 $y_{\perp}$            $\ge$ &  --  & 0.28    &  0.28  &  0.40   \\  \hline
\end{tabular} 
\caption{Values of the cuts for the purely hadronic selections
         which  are determined with the optimisation procedure
         described in Section~\protect\ref{sec:optimization}.\label{tab3}}

\end{center}
\end{table}

\begin{table} [htbp]
\begin{center} 
\begin{tabular}{|c|c|c|c|c|c|c|c|c|} \hline  
 & \multicolumn{2}{|c|}{Low \dm}& \multicolumn{2}{|c|}{Medium \dm}& 
\multicolumn{2}{|c|}{High \dm}& \multicolumn{2}{|c|}{Combined} \\  \hline
  & $N_{data}$ & $N_{exp}$ & $N_{data}$ & $N_{exp}$ & $N_{data}$ & $N_{exp}$ 
& $N_{data}$ & $N_{exp}$ \\  \hline
\chinop & 72 & 66.9 & 11 & 10.9 & 67 & 76.7 & 147 & 148. \\ \hline
\chinonn & 43 & 39.3 & 6 & 7.78 & 3 & 2.45 & 50 & 48.1 \\ \hline
\end{tabular} 
\caption[cascade]{Results for charginos and neutralinos:
           $N_{data}$ is the number of observed events and
           $N_{exp}$ is the number of expected events from Standard 
           Model processes for the total integrated
           luminosity collected at $\sqrt{s} = 189 \gev{}$.
\label{tab8}}
\end{center} 
\end{table} 
\begin{table} [htbp] 
\begin{center} 
\begin{tabular}{|c|r|r|r|r|} 
\hline  
{ }   &
            \multicolumn{2}{|c|}{~~LL~~} &
            \multicolumn{2}{|c|}{~~$\neutralino{1}\, \Wstar$~~} \\ \hline
{ \dm{} (\gev{}) }  & $\epsilon $ (\%) & $~N_{exp}$ &
                    $\epsilon $ (\%) & $N_{exp}$   \\
\hline 
3   &  1.6  & 20.3           &  1.9   & 39.4 \\ 
5   &  6.5   & 20.3          &  16.8  & 76.8  \\
10  &  19.2   & 27.7            &  8.5   & 2.9  \\
20  &  25.0 &   7.3           &  38.2  & 10.4  \\
30  &  30.0   & 7.3            &  46.6  & 7.5  \\ 
40  & 28.9   & 7.3            &  44.2  & 4.9  \\
50  & 26.9   & 7.3            &  25.3  & 4.9   \\ 
60  & 21.6   & 7.3            &  16.6  & 4.9  \\
75  & 34.5   & 34.6            &  20.8  & 55.5  \\
90  & 31.8   & 34.6            &  7.8   & 55.5  \\
\hline
\end{tabular} 
\caption[cascade]{Optimised chargino efficiencies ($\epsilon$)  
        for the purely leptonic (LL) 
        and for the
        $\neutralino{1}\,\Wstar$ decay mode.
        $N_{exp}$ is the number of events expected from  
        Standard Model processes.
        Results are given as a function of 
        $\dm$
        for $M_{\chargino{\pm}} = 94 \gev{}$ at $\sqrt{s} = 189 \gev{}$.
\label{tab4}}
\end{center}
\end{table}

\begin{table} [htbp]
\begin{center}
\begin{tabular}{|c|r|r|r|r|}
\hline  
{ }   &
            \multicolumn{2}{|c|}{~~LL~~} &
            \multicolumn{2}{|c|}{~~$\neutralino{1}\, \Zstar$~~} \\ \hline
{ \dm{} (\gev{}) }  & $\epsilon $ (\%)  &~$N_{exp}$ &
                    $\epsilon $ (\%) & $N_{exp}$ 
                    \\

\hline
6  & 9.1    & 9.5    & 3.5   & 35.9   \\
10  & 10.9  & 10.2   & 10.9  & 35.9    \\
20  & 27.3  & 4.2    & 9.2   & 3.4    \\
40  & 30.8  & 2.8    & 25.9  & 3.4    \\
60  & 34.1  & 19.2   & 35.2  & 7.7     \\
80  & 35.5  & 19.2   & 35.2  & 7.7    \\
100 & 29.3  & 21.4   & 20.7  & 7.7    \\
140 & 17.7  & 25.5   & 9.4  & 2.4    \\
180 & 10.5  & 25.5   & 8.7   & 2.4  \\
\hline 
\end{tabular} 
\caption[cascade]{Optimised neutralino efficiencies ($\epsilon$)
        for the purely leptonic (LL)
final states and for the $\neutralino{1}\, \Zstar$ decay mode. 
         Results are given as a function of         $\dm$
        for $M_{\neutralino{2}} + M_{\neutralino{1}} = 188 \gev{}$
at $\sqrt{s} = 189 \gev{}$.
 
\label{tab5b}}
\end{center} 
\end{table} 

\pagebreak 


%
%

\end{document}